\documentclass[a4paper,twocolumn,11pt,accepted=2022-09-19]{quantumarticle}
\pdfoutput=1
\usepackage[utf8]{inputenc}
\usepackage[english]{babel}
\usepackage[T1]{fontenc}
\usepackage{amsmath}
\usepackage{hyperref}

\usepackage{tikz}
\usepackage{lipsum}

\begin{document}

\title{Experimental entanglement generation for quantum key distribution beyond 1\,Gbit/s}

\author{Sebastian Philipp Neumann}
\email{sebastian.neumann@oeaw.ac.at}
\orcid{0000-0002-5968-5492}
\author{Mirela Selimovic}
\orcid{0000-0002-2291-997X}
\author{Martin Bohmann}
\orcid{0000-0003-3857-4555}
\author{Rupert Ursin}
\orcid{0000-0002-9403-269X}
\affiliation{Institute for Quantum Optics and Quantum Information, Boltzmanngasse 3, 1090 Vienna, Austria}
\affiliation{Vienna Center for Quantum Science and Technology, Boltzmanngasse 5, 1090 Vienna, Austria}
\maketitle

\begin{abstract}
 Top-performance sources of photonic entanglement are an indispensable resource for many applications in quantum communication, most notably quantum key distribution.
However, up to now, no source has been shown to simultaneously exhibit the high pair-creation rate, broad bandwidth, excellent state fidelity, and low intrinsic loss necessary for gigabit secure key rates.
In this work, we present for the first time a source of polarization-entangled photon pairs at telecommunication wavelengths that covers all these needs of real-world quantum-cryptographic applications, thus enabling unprecedented quantum-secure key rates of more than 1\,Gbit/s.
Our source is designed to optimally exploit state-of-the-art telecommunication equipment and detection systems.
Any technological improvement of the latter would result in an even higher rate without modification of the source.
We discuss the used wavelength-multiplexing approach, including its potential for multi-user quantum networks and its fundamental limitations.
Our source paves the way for high-speed quantum encryption approaching present-day internet bandwidth.
\end{abstract}

\section{\label{sec:Intro}Introduction}
Entanglement-based quantum key distribution (QKD) requires sources of photonic entanglement that exhibit high overall pair creation rates, high spectral brightness, low intrinsic loss, high fidelity to maximally entangled states and low maintenance.
These points are getting ever more important for achieving non-vanishing key rates over fiber and free-space quantum links which are governed by unavoidable strong losses.
In order to achieve this, different source designs have achieved remarkable individual figures of merit: Atzeni et al.\,\cite{Atzeni2018} achieved $2.2\times10^9$\,cps/mW overall brightness and Sun et al.\,\cite{Sun2019} $1.2\times10^9$\,cps/mW/nm spectral brightness, both in waveguide configurations. Liu et al.\,\cite{Liu2021} reported an average collection efficiency of $84.1$\% , Kaiser et al.\,\cite{Kaiser2014} as well as Joshi\,\cite{Joshi2014} showed 99.8\% polarization visibility, and the source of Tang et al.\,\cite{Tang2016a} even survived a rocket explosion.
For a recent comprehensive overview and a detailed discussion of the parameters in use, see Ref.s\,\cite{Anwar2021, Neumann2021b}.

While all of the reported sources show excellent merits regarding one or even two of these fundamental parameters, none of them exhibit outstanding overall performance necessary for first-grade real-world QKD applications.
In such applications, high \emph{overall brightness} is necessary to create high key rates required in telecommunication infrastructures today.
It is defined as the number of entangled photon pairs created in the source before all losses.
High \emph{spectral brightness}, i.e. the rate of photon pairs created per wavelength, enables efficient wavelength division multiplexing (WDM) of signals\,\cite{Aktas2016, Pseiner2021}, diminishes dispersion effects\,\cite{Neumann2021a} and will be necessary to couple to quantum memories in the future\,\cite{Heshami2016}.
\emph{Collection efficiency} is the probability of a photon created in the source being detected. High source-intrinsic collection efficiency allows to tolerate more (unavoidable) channel loss.
High \emph{visibility} of the entangled state allows the experimenter to efficiently perform error correction and privacy amplification in post-processing, which means that a larger fraction of the raw key can be utilized; additionally it can partly compensate for noise, detector jitter and channel loss.
\begin{figure*}[t!]
\includegraphics[width=\textwidth]{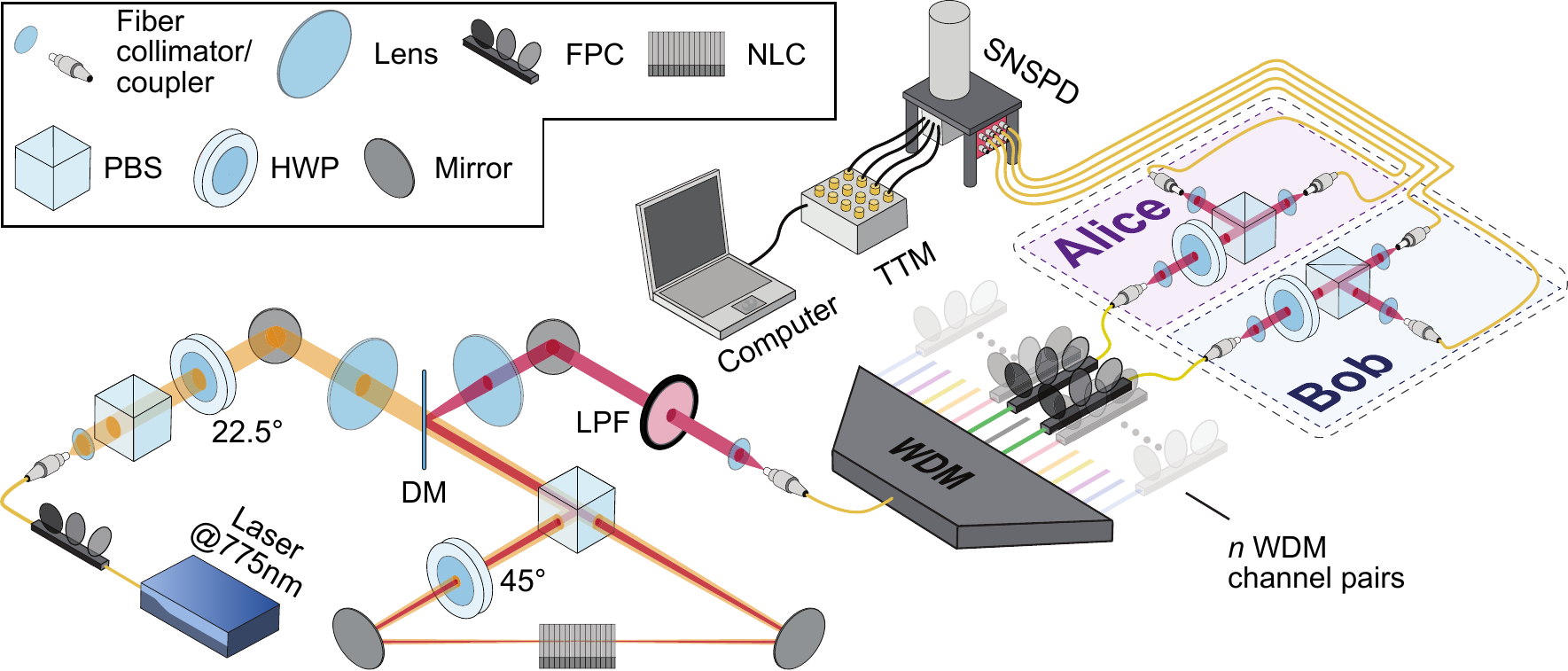}
\caption{\label{fig:source} Sketch of the setup. The  continuous-wave pump laser at 775.06\,nm is coupled into a single-mode fiber (SMF) and directed towards the bulk optics setup. After beam collimation, a polarizing beam splitter (PBS) and a half-wave plate (HWP) set the pump laser's polarization to $45^\circ$. A pump lens bidirectionally focuses the pump into the nonlinear crystal (NLC) placed inside a Sagnac loop. The loop consists of a dichroic PBS splitting the pump by transmitting (reflecting) the horizontally (vertically) polarized part. The reflected part passes a HWP at $45^\circ$. Thus, all pump photons are horizontally polarized when entering the NLC, where photon pairs are created via type-0 spontaneous parametric down-conversion (SPDC). They travel back to the PBS, where both propagation direction modes interfere and the SPDC photons' polarization states become entangled. Successively, a dichroic mirror (DM) separates the SPDC beam from the pump. The SPDC traverses a longpass filter (LPF) blocking residual pump light before being collected by a SMF. The broad total photon spectrum is subdivided by use of a WDM. For visibility measurements, polarization rotations in the fiber have to be compensated using fiber polarization controllers (FPC) in order to acquire the desired correlations. Measurements in mutually unbiased bases are realized by setting both receiver's HWP to either $0^\circ$ or $22.5^\circ$. The PBS output modes are coupled into SMF and directed to four channels of superconducting nanowire single-photon detectors (SNSPD), whose detection times are registered using a time-tagging module (TTM) and post-processed with a computer. For measurements of brightness and collection efficiency, the WDMs are directly connected to the SNSPDs. For the measurement of the total spectrum, the WDM was replaced by a 50:50 beam splitter, one arm of which was connected to a tunable filter (not depicted) before detection.}
\end{figure*}
The highest experimentally generated key rates as of today were acquired under laboratory conditions, without deployment of real-life links.
The record values are 10\,Mbit/s\,\cite{Yuan2018} in a decoy-state configuration, 26.2\,Mbit/s\,\cite{Islam2017} using time-bin qudits and 7.0\,Mbit/s\,\cite{Zhong2015} in an implementation with high-dimensional entanglement.
It should be noted that these values cannot be compared directly, since e.g. high-dimensional QKD has additional advantages such as higher noise-resilience\,\cite{Bulla2022arx}.

In this work, we present a source of polarization-entangled photon pairs performing competitively in all of the above-mentioned parameters, enabling unprecedented key rates beyond 1\,Gbit/s by exploiting polarization entanglement only.
The source was built in a bulk Sagnac-loop configuration deploying type-0 spontaneous parametric down-conversion (SPDC) inside a nonlinear crystal producing polarization-entangled photon pairs at telecom wavelength.
Using bulk polarization measurement modules and a tunable blazed-grating filter, respectively, we quantified brightness, collection efficiency, spectral bandwidth and polarization visibility for different WDM channels as well as the full spectrum.
We find that using 66 channel pairs of off-the-shelf WDM devices, the source could supply a total of 1.2\,Gbit/s secure key in a point-to-point configuration.
Even higher values of up to 3.6\,Gbit/s are conceivable when using narrow ultra-dense WDM channels and pumping
with high laser power.
Additionally, we show that by using the full $106\,$nm-bandwidth spectrum of our source, a fully connected local quantum network with up to 33 users could be created.
Our results provide an essential contribution towards high-key-rate quantum communication necessary for future quantum infrastructures.

\section{\label{sec:Results}Results}
\subsection{Source design}
To arrive at key rates above $1\,$Gbit/s, we need to determine the outstanding values of the source's spectral bandwidth, collection efficiency, brightness and visibility values, which we will set forth in the following.
The experimental set-up is depicted in Fig. \ref{fig:source}.
The source was built in a bulk Sagnac configuration with the loop containing a type-0 temperature-stabilized nonlinear crystal (NLC).
It was bidirectionally pumped using a narrow-band 775.06\,nm continuous-wave laser with its focusing parameters optimized for high-brightness SPDC and a maximum power of 500\,mW.
It produces telecom-wavelength entangled photon pairs with their spectrum centered around 1550.12\,nm.
Carefully chosen collimation and coupling optics allow for the SPDC's low-loss insertion into a single-mode fiber.
For a more detailed description of the source's working principle, trade-offs and design considerations, see the caption of Fig. \ref{fig:source} and the Methods section.
The photons were detected by use of two fiber-coupled superconducting nanowire single-photon detector (SNSPD) channels connected to a time-tagging module (TTM).
From its time stamps, our computer software calculated $g^{(2)}$ correlations between the channels, from which we identified the entangled photon pairs.

\subsection{Evaluation of source performance parameters}
\begin{figure}
\includegraphics[width=\columnwidth]{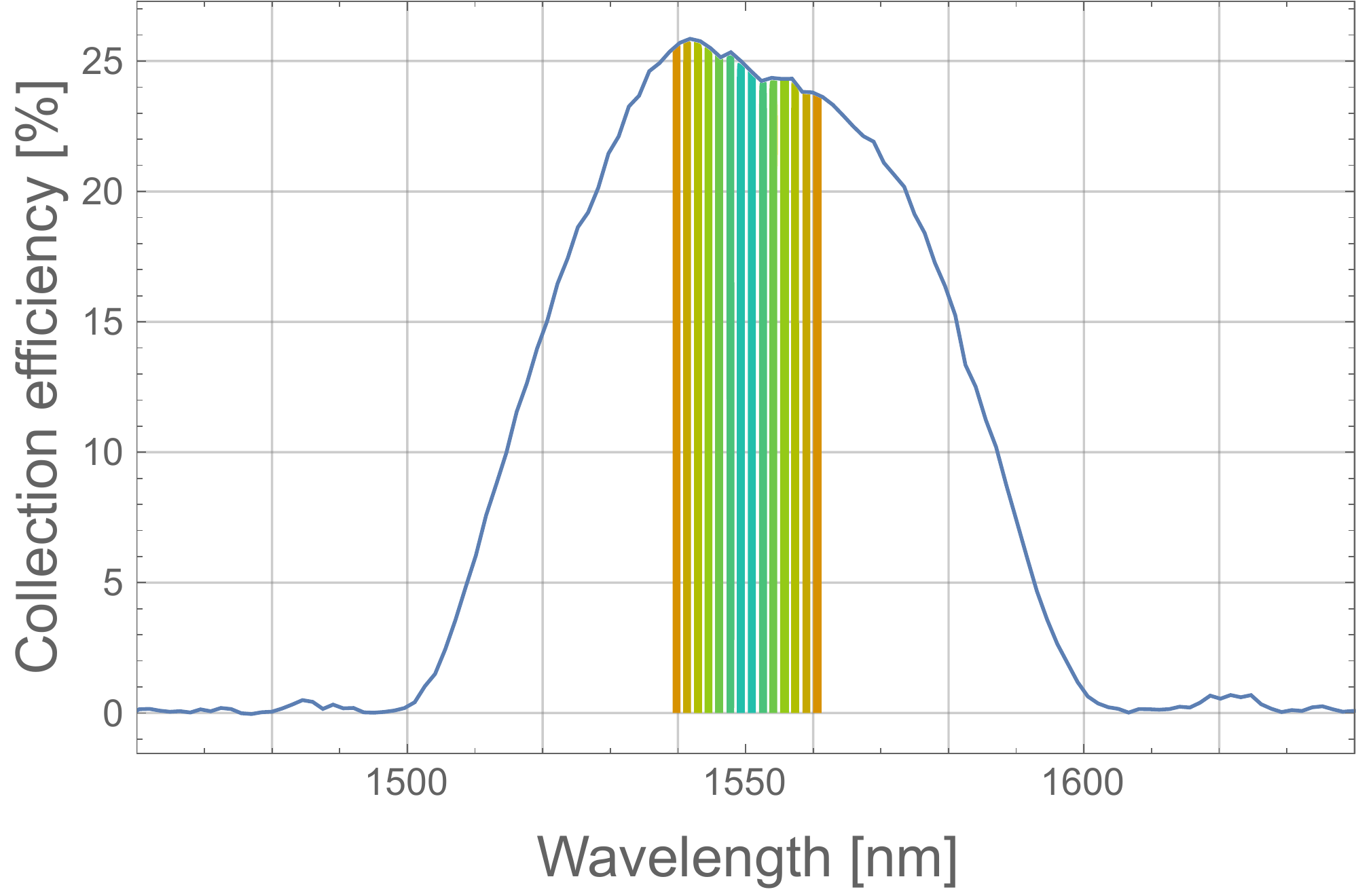}
\caption{\label{fig:spectrum} Measured collection efficiency of the source per wavelength. The measurement was carried out using a tunable wavelength filter based on a rotatable blaze grating, which one photon of a pair passed. The graph was obtained by determining the ratio of coincidence counts vs. singles of the lossy filter arm and normalizing to the highest collection efficiency obtained in the WDM measurements (see Fig. \ref{fig:heralding}). WDM channel pairs carrying entangled photon pairs are shown as slices of the same color. The slight asymmetry of the spectrum can be explained by varying coupling efficiencies of the tunable filter and the single-photon detectors.}
\end{figure}

Figure \ref{fig:spectrum} shows the source's collection efficiency over the full SPDC spectrum.
The graph was acquired by probabilistically separating the entangled pairs with a 50:50 in-fiber beam splitter before detection and using a free-space grating-based tunable wavelength filter in one of its arms.
The spectrally resolved collection efficiency is required to quantify the portion of the spectrum usable in QKD.
The source spectrum can be optimally exploited for QKD by deterministically separating entangled photon pairs using WDM channels.
Due to energy conservation during the SPDC process, entangled photons are found in channel pairs equidistant from the spectrum's central wavelength. Such pairs are depicted in the same colors in Fig. \ref{fig:spectrum}.
Each of the $n$ channel pairs can be considered an independent carrier of photonic entanglement\,\cite{Wengerowsky2018a}.

\begin{figure}
\includegraphics[width=\columnwidth]{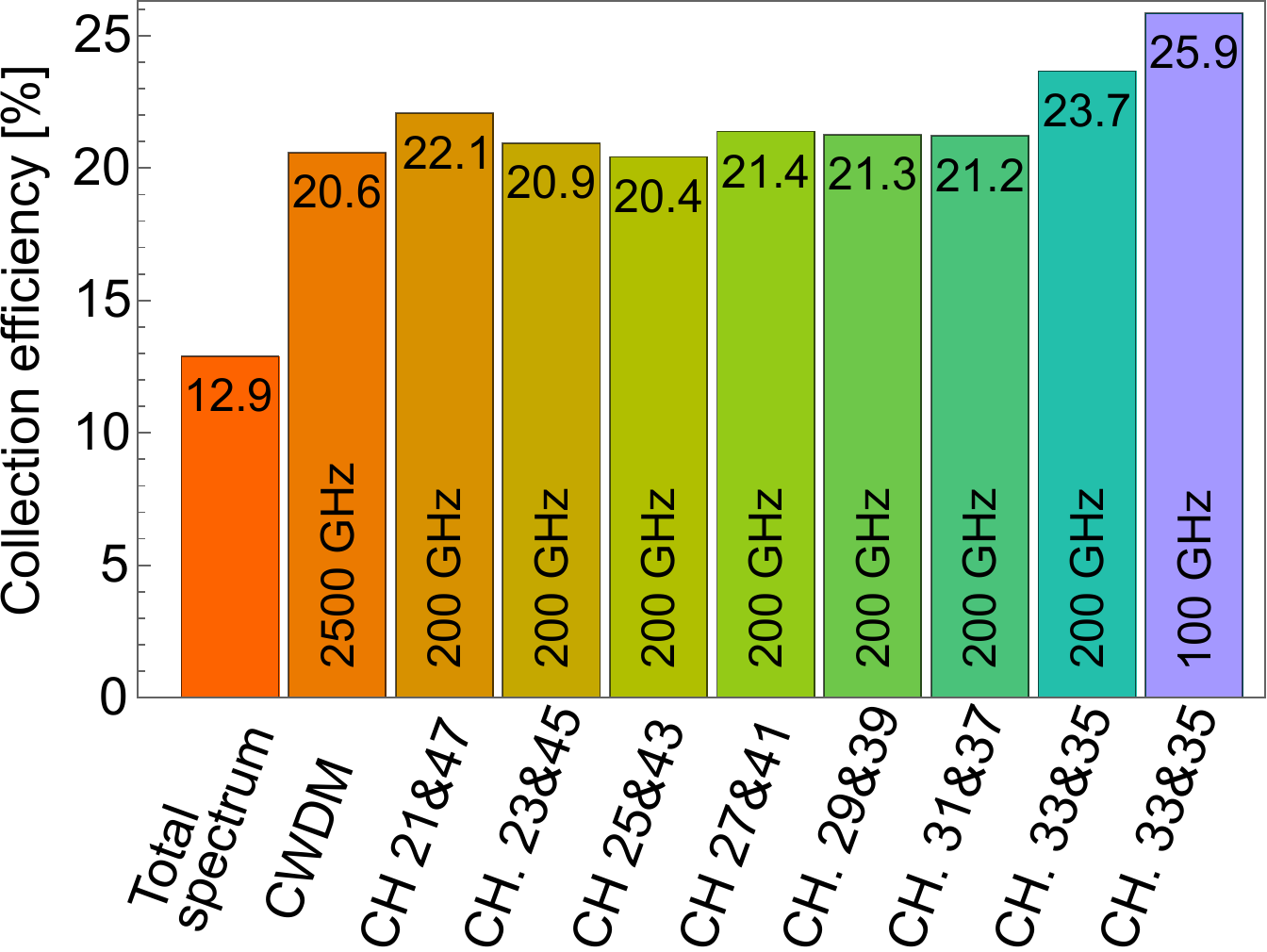}
\caption{\label{fig:heralding} Average measured collection efficiencies per wavelength-channel pair obtained by determining the ratio between coincident and single detector counts. Note that the lower values for the full spectrum originate in part from the fact that a probabilistic beam splitter was used to separate the photons, and in part from the fact that far from the central wavelength, the source intrinsically shows lower collection efficiencies due to its focusing parameters \,\cite{Bennink2010}.}
\end{figure}

To precisely determine collection efficiencies and brightness values, matching WDM channel pairs were connected directly to the SNSPDs.
To ensure straightforward comparability with other source designs, we did not subject the photon pairs to long-distance link attenuation.
The single-mode fibers in use added up to no more than 10\,m length.
All of the following collection efficiencies include coupling and transmission losses of WDMs and fibers, leakage from imperfect wavelength channel separation and SNSPD detection efficiencies.
We define the collection efficiency $\eta$, sometimes called ``heralding" or ``Klyshko"\,\cite{Klyshko1980} efficiency, as $\eta=CC/\sqrt{S_A\cdot S_B}$, where $CC$ are the coincident counts between the communicating partners' detectors with single count rates $S_A$ and $S_B$.
Figure \ref{fig:heralding} shows collection efficiency values for different standard WDM channels of $100$ and $200$\,GHz (dense WDM) and 2500\,GHz (coarse WDM, CWDM).
Collection efficiencies stay above 20\% on average in a $56.3$\,nm range around the central SPDC wavelength.
As a comparison, averaging over the full spectral range, the value decreases to $12.9$\%.
This value was acquired using a 50:50 in-fiber beam splitter and is in accordance with Fig. \ref{fig:spectrum}, since coupling into the tunable filter's collecting single-mode fiber is less efficient far from the central wavelength due to chromatic aberration of the coupling optics and wavelength-depended mode structures.

\begin{figure}
\includegraphics[width=\columnwidth]{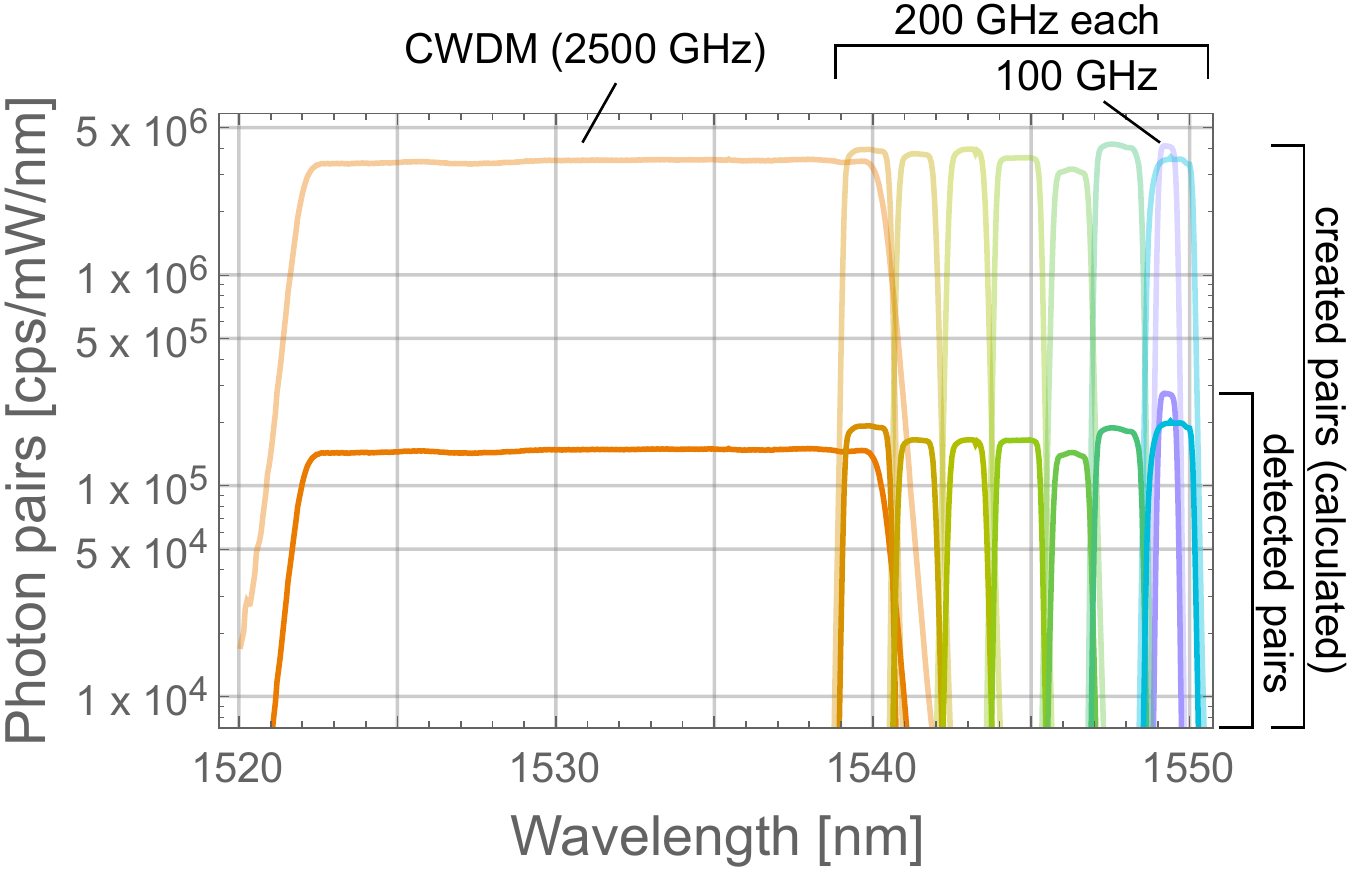}
\caption{\label{fig:brightness} Source brightness per wavelength. Brightness values (transparent) and measured coincidences (solid) of the source for measured WDM channel pairs per mW. The central wavelength is $1550.12$\,nm, and only the channels with the lower wavelength of each channel pair are depicted for simplicity.
Brightness values were calculated from measured coincidences and collection efficiencies.}
\end{figure}

Figure \ref{fig:brightness} shows spectral brightness values, i.e., the number of photon pairs per pump power, wavelength and time, for the same WDM channels as used for Fig. \ref{fig:heralding}.
Solid lines refer to \emph{detected} pair rates, while faint lines are calculated pair creation rates in the crystal before any loss. This parameter is called spectral brightness $B$.
The latter depends on measured pair rates and collection efficiencies as $B=CC/\eta^2$.
The highest spectral brightness value of $B_{31+37}=4.17\times10^6$\,cps/mW/nm was found in the 200\,GHz WDM channel pair 31+37.

\begin{figure}
\includegraphics[width=\columnwidth]{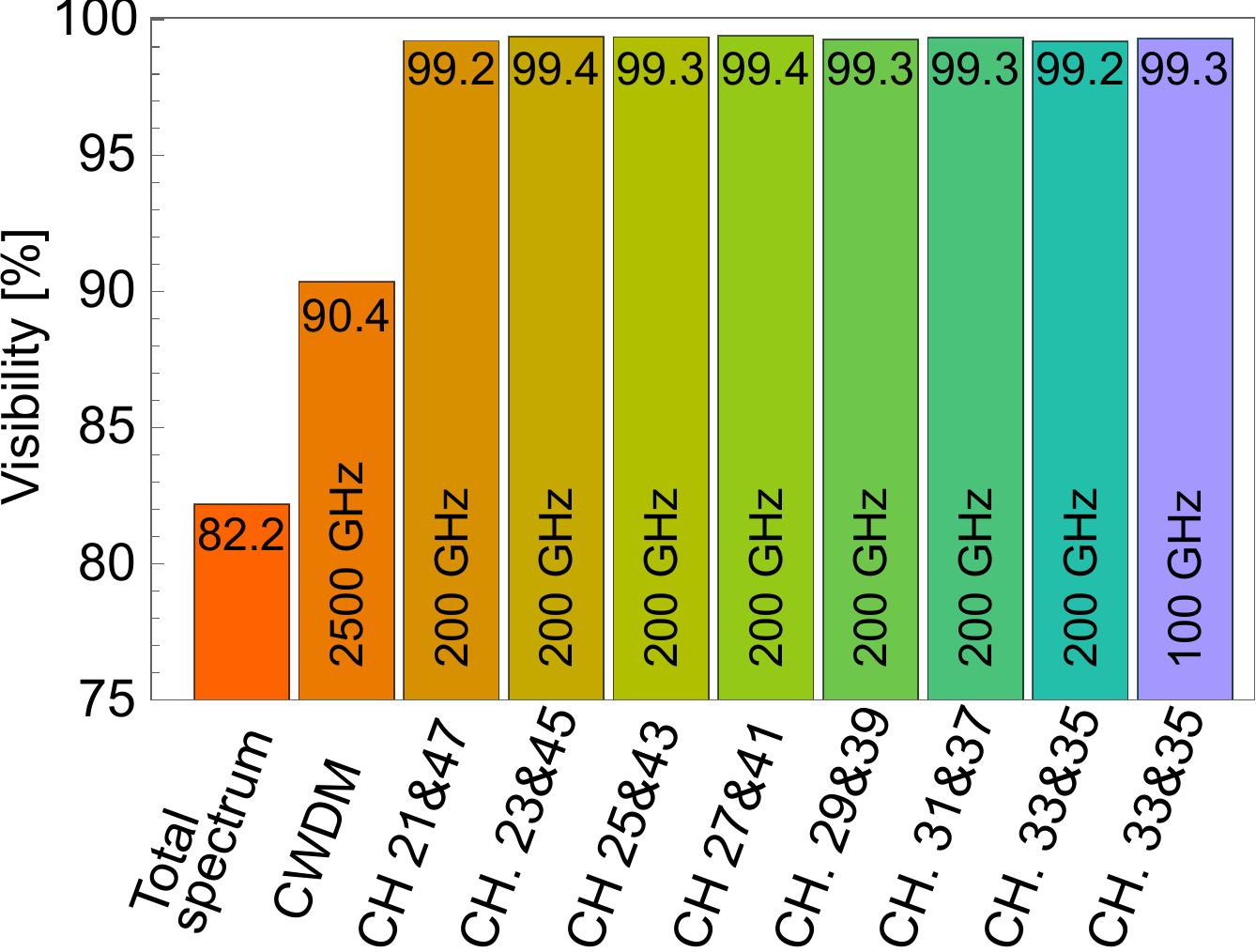}
\caption{\label{fig:vis} Visibility per wavelength-channel pair. Average values of visibility $V$ obtained by measuring correlations in polarization in two mutually unbiased bases. We measured $V>99.2$\% for all WDM spectral widths used for key calculations. We attribute lower visibility values for the 2500\,GHz coarse WDM (CWDM) and the full spectrum to  wavelength-dependent polarization rotations in the fibers, which can be compensated individually if subdividing in narrow spectra.}
\end{figure}

Figure \ref{fig:vis} demonstrates our source's exceptional fidelity to a maximally entangled state with measured polarization visibilities $V$ of up to 99.4\% in two mutually unbiased bases.
To arrive at these values, bulk polarizing beam splitters with single-mode coupled output ports were implemented between WDMs and SNSPDs (cf. Fig. \ref{fig:source}).
$V$ stayed above $99.2$\% for all observed $100$ and $200$\,GHz channels.
Since polarization rotations in fiber are wavelength dependent, no full polarization compensation using fiber polarization controllers (FPC) can be achieved for broad spectra.
However, even for the broader CWDM channels and the full spectrum, the quantum bit error rate (QBER) $E=(100\%-V)/2$ stays above the 11\% limit necessary for secure key creation\,\cite{Shor2000}.

As a final figure of merit, we want to point out our source's stability:
All of the above data was taken more than 6 months after source alignment, with no certifiable performance degradation during this period.
The only active stabilization necessary was carried out by an electronically controlled oven restricting crystal temperature fluctuations to $<0.01^\circ$\,C, while no performance degradation could be observed for changes $<0.1^\circ$\,C.

\subsection{Secure key rate analysis}
\begin{figure}[t!]
\includegraphics[width=\columnwidth]{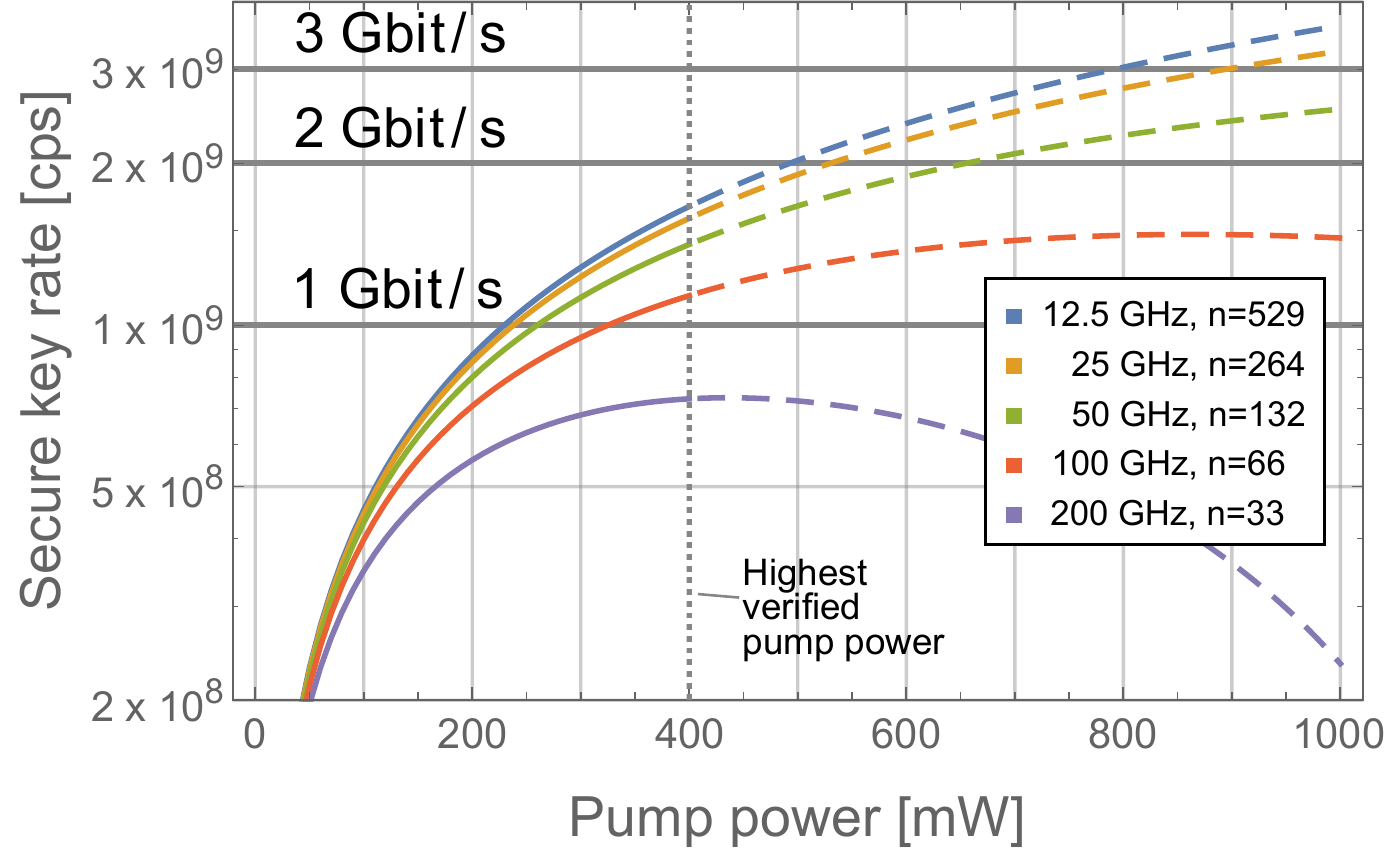}
\caption{\label{fig:key} Quantum-secure key rates per pump power.
Expected key rates per pump power at $775.06$\,nm for different wavelength demultiplexing scenarios after sifting and error correction, taking into account multipair statistics. The WDM channels under consideration follow standards defined by the International Telecommunication Union and we indicate the number of used channel pairs $n$. Already for 100\,GHz channels, our source could provide key rates above 1\,Gbit/s. While we experimentally verified source operation without performance decrease up until 400 mW pump power, it is reasonable to assume that 1000 mW are still feasible\,\cite{HCphotonics}. This could allow for more than 2.0\,Gbit/s in a 50\,GHz demultiplexing scheme, while 25\,GHz and 12.5\,GHz enable more than 3.0\,Gbit/s. The relative increase in key rate for ever narrower channels becomes smaller due to the accompanying increase of the entangled photons' coherence time.}
\end{figure}

From this experimental data, maximum secure key rates can be inferred according to the model in Ref.\,\cite{Neumann2021b}. This model considers the probabilistic multi-photon-pair statistics of a continuous-wave-pumped entangled-photon source.
An in-depth analysis of these statistics is necessary, since simply increasing the pump power can be detrimental to the key rate.
This is because multi-photon pairs lead to so-called accidental coincidences and thus an increased QBER, making it necessary to use a larger portion of the key for classical post-processing. Therefore, there exists an optimal pump power.
This optimum additionally depends on the wavelength-channel width, leading to a trade-off between the rate of detected photon pairs and the accidental detection probability per wavelength-channel pair.
The source can be operated as is, by optimizing the laser power inside the crystal.
Using the collection efficiency, brightness, and visibility values experimentally achieved in our experiment, we simulate QKD implementations with different WDM scenarios in Fig. \ref{fig:key}.

Here, each WDM-channel pair can provide a secure key rate $R^s_k$ depending on its individual collection efficiency and entanglement quality.
In this work, we are concerned with the total key rate $R^s=\sum_{k=1}^n R^s_{k}$ achievable with $n$ channel pairs from our source.
Our calculations, depicted in Fig. \ref{fig:key}, show that already with standard off-the-shelf 100\,GHz WDM channels, 1.2\,Gbit/s secure key rate could be achieved at 400\,mW pump power when deploying suitable high-end detectors (see Methods for more details on assumptions about detectors).
With 132 channels of 50\,GHz width, $R^s=2.0$\,Gbit/s can be achieved with 660\,mW pump power.
25\,GHz channels would even allow for $3.0$\,Gbit/s at 900\,mW, and reducing the spacing further to 529 channel pairs of 12.5\,GHz, the same key rate value could be reached already with 800\,mW pump power.
When pumping with 1000\,mW in the latter WDM configuration, a maximum value of 3.6\,Gbit/s is possible.

In a short-distance and low-loss approximation, additional transmission channel losses occurring in any real-life QKD scenario can simply be multiplied with the optimized key rates.
As an example, a 10\,km fiber would add 2\,dB loss and therefore lower the maximum key rates to about 63\% of the values depicted in Fig.\,\ref{fig:key}.
This would in turn require the deployment of 25\,GHz channel spacing while still pumping with 400\,mW to reach the 1\,Gbit/s limit, or 50\,GHz spacing and 470\,mW pump.
Additionally, a successful QKD implementation has been carried out over a record-breaking distance of 248 km with 79 dB loss, using an adapted version of this work’s source\,\cite{Neumann2022arx}.
Please see the Methods section for details on key rate formulas and optimization.

\section{Discussion}
We have presented a stable source of polarization-entangled photon pairs with high total and spectral brightness, high collection/heralding efficiency and extremely high state fidelity. 
Calculating the quantum secure key rates that our source could sustain when operating with sufficiently performing single-photon detectors, we arrive at key rates above 1\,Gbit/s with off-the-shelf wavelength-division-multiplexing devices and laser powers below the crystal's damage threshold.
When relaxing the latter requirements, key rates of more than 3\,Gbit/s are conceivable with our state-of-the-art source:

Firstly, stronger pump laser power could increase the effective spectral and overall brightness.
To mitigate the risk of damage to our setup, we restricted ourselves to powers of no more than 400\,mW, for which we could still certify undiminished source performance.
However, if one were to install laminar airflow boxes to keep dust away from the optical surfaces, specifications by the crystal manufacturer suggest that powers of 1000\,mW and beyond are feasible\,\cite{HCphotonics}.
Secondly, using narrower WDM spacings and therefore higher channel pair numbers $n$ reduces the number of undesired accidental correlation measurements and therefore enhances the key rate\,\cite{Pseiner2021}.
Deploying WDMs narrower than 100\,GHz might require customization, but is possible in principle and also covered by ITU standards\,\cite{ITUgrid2020}.
However, going below 6.25\,GHz is hardly beneficial anymore.
This is because the entangled photons' coherence time is in the order of tens of picoseconds in this case, which deteriorates the timing precision necessary for photon pair correlation.
Ultimately, even assuming perfect temporal photon detection, this time-bandwidth product effect represents the physical limit of increasing the rate of polarization and time-bin based QKD protocols.
As a side remark, further narrowing the channel width might nevertheless be beneficial in order to address quantum memories for future quantum computing or quantum repeater schemes\,\footnote{Despite efforts to create quantum memories at terahertz bandwidths\,\cite{England2013}, most current quantum memories exhibit bandwidths of 5\,GHz and below\,\cite{Heshami2016}.}.

In a real-world QKD implementation, where many channels including their respective detection system can terminate at one and the same communication partner, $n$ channel pairs can connect between $2$ and $2n$ users individually.
However, due to its broad spectrum, our source is also ideally suited for fully connected multi-user quantum network configurations\,\cite{Wengerowsky2018a}.
With the 66 channel pairs available when using 100\,GHz spacing, our source could fully connect 12 users in a trusted-node-free network design without any probabilistic multiplexing\,\cite{Joshi2020}.
Deploying 529 channel pairs of 12.5\,GHz width, this number increases to 33 fully connected users.

Although our source is ideally suited for a large variety of applications, its practical deployment is limited by current single-photon detector performance.
For calculating the overall key rate, we assume our detector's quantum efficiency, specified to be $80$\% by the manufacturer, to stay constant for all wavelength channels and count rates.
We assume 38\,ps jitter of the full detection system including time-tagging electronics, which was the lowest value we observed during our experiment.
Most importantly, we have simultaneously assumed maximum detector count rates of 200\,MHz.
While such count rates can be achieved in state-of-the-art experiments\,\cite{Perrenoud2020}, so far there exists no detection system simultaneously exhibiting high detection efficiency as well as low jitter.
We note, however, that in (high-loss) long-distant communication scenarios, maximum count-rate limitations do not pose any practical problems for QKD implementations due to the reduced number of registered photons.
But even in these cases, low jitter is crucial to avoid accidental two-fold clicks of uncorrelated photons.
Thus, it becomes apparent that although recent research shows promising approaches\,\cite{Rambo2021arx, Korzh2020,Perrenoud2020}, detector technology has yet to catch up with high-end entangled-photon-pair sources such as the one presented in this work.

We want to stress that for the claims presented in this work, no problems or challenges of the source design were shifted to the detection devices.
There is no conceivable enhancement to the source that could lead to a QKD performance increase without a significant prior advance in detector technology.
As of today, detectors are the limiting factor for achieving high key rates: temporal jitter as well as dead time of SNSPDs (and, even more so, of semiconductor-based single-photon detectors) can neither resolve nor register the extraordinarily high pair creation rates of our high-end source.

\section{Conclusion}
We have described a source capable of providing more than $1\,$Gbit/s secure key rate using off-the-shelf components.
To the best of our knowledge, this is the brightest source with simultaneously optimized collection efficiency and visibility up to date.
Exceptional visibility of the source's polarization-entangled photon pairs enables quantum bit error rates below $0.4\,$\%.
Measured collection efficiencies of up $25.9$\% provide high photon yield.
Damage thresholds measurements by the crystal manufacturer\,\cite{HCphotonics} suggest that the crystal could be pumped with up to 1\,W, which would create more than $10^{11}$ photon pairs per second over the full spectrum.
These exceptional entangled photon pair creation rates cannot be resolved by single-photon detection systems as of today.
In order to fully exploit the potential of readily available entangled photon sources such as the one presented in this work, detector research has to aim towards simultaneously lowering both dead time and temporal detection imprecision of single-photon detectors.
Thus, we have identified the most pressing problem in current QKD technology as the trade-off between maximum count rate and timing jitter of modern detection systems, which limits the performance of state-of-the-art source technology.
\section*{Acknowledgments}
We acknowledge European Union’s Horizon 2020 programme grant agreement No. 857156 (OpenQKD) and the Austrian Academy of Sciences.
\newpage
\section{Methods}
\subsection{\label{sec:sourcedesign}Source design and working principle}
    The source is a bulk optics telecom-wavelength source making use of SPDC inside a nonlinear periodically poled 5\% magnesium-doped congruent lithium niobate crystal with type-0 phase-matching\,\cite{Fejer1992}.
    The crystal is placed inside a Sagnac loop to enable coherent superposition of orthogonally polarized SPDC modes.
    Active temperature stabilization to $\pm0.01^\circ$ ensured optimal phase-matching inside the crystal.
    The crystal is bidirectionally pumped using a 775.06\,nm narrow-band ($\approx50$\,kHz linewidth) continuous-wave laser.
    The loop was built as small as possible to allow for strong focusing of the pump, which has to be carried out by a single lens outside the loop in order to be able to separately optimize pump focusing and SPDC collimation.
    Additionally, using one pump lens benefits indistinguishability of the beam profile in both of the loop's propagation direction modes.
    The focusing parameter $\xi$ was chosen carefully according to efficiency considerations in Ref.\,\cite{Bennink2010}.
    We put emphasis on obtaining a large ratio of crystal length to Rayleigh length of the pump beam in order to obtain high pair creation rates.
    Concretely, this corresponded to a focus length $f=254$\,mm for the pump lens and $f=200$\,mm for the SPDC collection lens.
    To ensure good fidelity of the pump beam to a TE$_\textrm{00}$ mode, an aspheric lens with $f=18.4$\,mm was used to couple out of the single-mode fiber.
    The comparably strong pump focusing required for this goal can lead to divergence-induced degradation of the PBS extinction ratio, which negatively influences the brightness.
    We therefore regularly checked the PBS's performance during source construction with lasers at both pump and SPDC wavelength.
    The loop length is restricted on one hand by the focal length of the lenses, since tighter focusing corresponds to higher brightness.
    On the other hand, it is limited by the length of the crystal including its temperature control, which has to fit in the long side of the loop.
    Additionally, the pump and SPDC beams must not be clipped when entering and leaving the crystal, therefore again limiting beam divergence.
    To account for these different trade-offs, we chose a loop length of $\approx35$\,cm to house a nonlinear crystal of 50\,mm length for enabling the SPDC process.
    In combination with the pump beam's calculated Gaussian profile, this resulted in $\xi=1.99$.
    The SPDC beam's focus parameters were matched to the same value as the pump's, which required different collimation and collection parameters due to their different wavelengths.
    The entangled photons were coupled into an SMF with a 1550\,nm anti-reflection coating via an aspheric lens.
    After passing a longpass filter, all SPDC photons were coupled into one standard SMF-28 single-mode fiber.
    To avoid laser-induced damage to any of the optical components, especially by dust burnt to their surfaces, the source was covered by a laminar air flow chamber, thus avoiding contamination of the set-up.
    For a detailed description of the source's working principle, we additionally refer the reader to Fig. \ref{fig:source}.\\
    To ensure outstanding performance of our source, not only the correct choice of parameters, but also an elaborate alignment procedure was essential.
    To this end, we measured the power of light being reflected back to the pump laser's isolator to ensure perfect alignment of the loop.
    This also allows for perfect collimation of the pump beam, which can be guaranteed by maximizing the back-coupling.
    Additionally, one alignment step was to couple polarization-controlled 1550.12\,nm laser light amplified to $>200$\,mW  by use of an erbium-doped fiber-amplifier out of the SPDC collection fiber to make use of up-conversion for reversing the beam paths and aligning the full set-up with strong laser light.\\
    Separation of the entangled photon pairs was carried out using standard dense wavelength-division multiplexing (DWDM) modules with a channel spacing of 200 as well as 100\,GHZ according to the ITU grid.
    We want to emphasize that we deployed off-the-shelf telecommunication devices for this task. Their spectrum's full width at half maximum (FWHM) amounts to only about $75\%$ of the channel spacing.
    Deploying custom-made DWDM channels with steeper edges, thus allowing broader FWHM, could therefore increase the usable part of the spectrum by up to $25\%$.
    The WDM channels carrying the respective entangled photons of a pair were connected to two channels of a SNSPD system with 80\% detection efficiency according to the manufacturer.
    Detection events were assigned a time stamp with 1\,ps bin width by use of a time-tagging module.
    From two-fold coincident counts between the two detector channels, we calculated a $g^{(2)}$ correlation function for each channel pair.
    The FWHM of the correlation peak amounted to $38$\,ps, which is equivalent to the total timing jitter of both detection systems.

\subsection{\label{sec:Results2}Benchmarking source performance}
To determine the source brightness, i.e. the number of entangled pairs produced inside the crystal via SPDC before any losses per second, the WDM channels were connected to the SNSPDs directly.
From their collection efficiencies\,\cite{Klyshko1980}, one can infer the total channel losses and thus the pair production rates (see Fig. \ref{fig:brightness} and the related discussion).
Measurements were carried out for pump powers of $50$\,\textmu W to keep the ratio of accidental coincidence counts to pair counts low.
This is important in order to neither overestimate the heralding efficiency nor underestimate brightness due to uncorrelated accidental counts mistakenly registered as coincidences.
Only in that case, it is admissible to calculate the brightness as $B=CC/\eta^2=S_AS_B/CC$, where noise counts have been subtracted from single count rates.
Additionally, to check for crystal damages and overall source performance in high-power regimes, we exemplarily checked $B_{33+35}^\textrm{100\,GHz}$, the brightness for the 100\,GHz channel pair $33+35$, for a pump power of 400\,mW over high-loss fiber links with $\approx40$\,dB attenuation each.
We confirmed $B_{33+35}^\textrm{100\,GHz}=4.10\times10^6$\,cps/mW/nm for both 50\,\textmu W and 400\,mW, therefore verifying that for powers up to 400 mW, as required for our claim of $1.2$\,Gbit/s secure key rate, no gray-tracking, crystal damages or any other decreases of source performance occur. However, we chose not to increase the power even further in order not to risk compromising other experiments the source is needed for, although we are confident that even higher power levels are feasible\,\cite{HCphotonics}.

To determine the source's collection efficiency per wavelength over its full spectral range, the WDMs were replaced by a 50:50 fiber beam splitter (FBS).
One FBS output port was connected to a SNSPD channel directly, while the other one was directed to a free-space rotatable blazed grating reflecting the incoming signal with 1.46\,nm/mrad angular dispersion towards an SMF.
A 1.25 nm (FWHM) wide portion of the signal was coupled into the SMF, which effectively acted as a wavelength filter.
The SMF was connected to a second SNSPD channel.
We determined the ratio between coincident counts of both channels and the single counts of the filter channel for different angle settings of the channel.
All measurements were obtained using the same detectors and thus include the SNSPD's wavelength dependency.
To account for excess loss due to inefficient coupling and dead-time loss in the first SNSPDs, we normalized for the collection efficiency achieved with the 100\,GHz WDMs and thus arrived at Fig. \ref{fig:spectrum}.
These values include all losses from creation of the photon inside the crystal to its detection.
To evaluate the source's intrinsic collection efficiency, one must subtract all non-source-specific losses emerging from fiber connections, WDMs and detection efficiency.
We quantify these losses, using power measurements with classical lasers and SNSPD manufacturer specifications, resulting in a transmission factor of 57\%. 
This is in reasonably good agreement with the 25.9\% maximum heralding efficiency measured for our source.
To account for the remaining discrepancy, one has to consider imperfect and lossy optics in the source as well as manufacturing tolerances for the pump- and SPDC-fiber mode-field diameters.
Even more importantly, there exists a fundamental collection efficiency limit of $\approx$75\% for $\xi=1.99$ at degenerate wavelength\,\cite{Bennink2010}.

The entangled photons' state fidelity was measured using two polarization-detection modules with two detector channels each (see Fig. \ref{fig:source}).
This way, erroneous counts could be quantified directly.
The error was determined in two mutually unbiased bases which were set using half-wave plates (HWP).
Fidelity measurements were carried out for 7 channel pairs with 200\,GHz width and resulted in $>99.2$\% visibility, i.e. $<0.4$\% QBER for all channel pairs.
These values were determined with pump powers low enough to ignore noise-induced coincidence counts, which occured with probabilities of less than $10^{-4}$ per registered photon pair.
Deploying broader wavelength channels generally leads to lower visibilities, since in-fiber polarization rotations along the WDMs and SMFs leading to the detectors are wavelength-dependent.
Thus, compensation using FPC cannot be carried out equally well for the full channel spectrum.
Therefore, for the 18.4\,nm coarse WDM, the average visibility only reached $90.4$\%. 
This problem can easily be mitigated by deploying more WDM channels with denser spacing and individual polarization compensation.

\subsection{\label{sec:key}Secure key rate calculations}

To arrive at final secure key rates, we made use of the model presented in Ref.\,\cite{Neumann2021b}, which was verified with the very same source as the one presented in this work.
In the following, we shortly outline the formulas used to arrive at the key rates depicted in Fig. \ref{fig:key} and refer to\,\cite{Neumann2021b} for in-depth explanations:
\begin{widetext}
\begin{align}
    \textrm{CC}^\textrm{t}&=\:B_\textrm{tot}(P,\Delta\lambda)\:\eta(\lambda_0-n\Delta\lambda,\Delta\lambda)\:\eta(\lambda_0+n\Delta\lambda,\Delta\lambda)\:\textrm{erf}\Bigg[\sqrt{\textrm{ln}2}\:\frac{t_\textrm{CC}}{\sqrt{t_\Delta^2+\sigma_\textrm{C}^2(\lambda_0,\Delta\lambda)}}\Bigg],\label{eq:line1}\\
    \textrm{CC}^\textrm{acc}&=\Big(B_\textrm{tot}(P,\Delta\lambda)\:\eta(\lambda_0-n\Delta\lambda,\Delta\lambda)+2\textrm{DC}\Big)\Big(B_\textrm{tot}(P,\Delta\lambda)\:\eta(\lambda_0+n\Delta\lambda,\Delta\lambda)+2\textrm{DC}\Big)\:t_\textrm{CC}.\label{eq:line2}
\end{align}
\end{widetext}
$\textrm{CC}^\textrm{t}$ are the so-called ``true" coincidences per wavelength channel, i.e. those entangled photons that actually yield a click, where $B_\textrm{tot}=B\cdot P\cdot\Delta\lambda$ are the photons produced in the crystal at pump power $P$ over WDM width $\Delta\lambda$.
For the sake of consistency, we used $B=B_{33+35}^\textrm{100\,GHz}=4.10\times10^6$\,cps/mW/nm of the same 100\,GHz channel pair that was used both to check source performance for high laser powers and to normalize the experimentally determined wavelength-dependent collection efficiency $\Lambda(\lambda)$ shown in Fig.\,\ref{fig:spectrum}.
The latter includes coupling and focus-mismatch losses of the source, attenuation and leakage in the WDM devices and fibers leading to the SNSPDs, and detection efficiencies of the latter.
It relates to the heralding efficiency as
\begin{align}
\eta(\lambda,\Delta\lambda)=\frac{0.75}{\Delta\lambda}\int\limits_{\lambda-\frac{\Delta\lambda}{2}}^{\lambda+\frac{\Delta\lambda}{2}} \Lambda(\lambda)d\lambda,
\end{align}
where 0.75 is the spectral fill-factor of the WDM channels. $\lambda_0=1550.12$\,nm is the central wavelength of the SPDC spectrum. The error function erf describes how, depending on the length of the coincidence window $t_\textrm{CC}$, some photons of $\textrm{CC}^\textrm{t}$ are lost due to their Gaussian-shaped temporal distribution, which is in turn defined by detector jitter $t_\Delta=38$\,ps and photon coherence time $\sigma_\textrm{C}(\lambda_0,\Delta\lambda)$, which we approximated from central wavelength and respective WDM width. The accidental coincidences $\textrm{CC}^\textrm{acc}$ depend on the single count rates of each communicating partner, including the dark count rates per detector DC, and on $t_\textrm{CC}$. 
From these quantities, one can calculate the key rate $R_s$ of Fig.\,\ref{fig:key} in the asymptotic limit depending on pump power and wavelength channel spacing:
\begin{widetext}
\begin{align}
    R_s(P,\Delta\lambda)&=\sum_{(\lambda_A,\lambda_B)}^n\max_{t_\textrm{CC}}\Bigg((\textrm{CC}^\textrm{t}+\textrm{CC}^\textrm{acc})
    \Big(1-2H_2\Big[\frac{
     \textrm{CC}^\textrm{t}\:e^\textrm{pol}+\frac{1}{2}\textrm{CC}^\textrm{acc}
    }{\textrm{CC}^\textrm{t}+\textrm{CC}^\textrm{acc}}\Big]\Big)\Bigg).\label{eq:line3}
\end{align}
\end{widetext}
Here, we sum over wavelength pairs $\lambda_A$ and $\lambda_B$, where $(\lambda_A+\lambda_B)/2=1550.12$\,nm holds for each of the $n$ pairs.
The binary entropy function $H_2=-x\log_2(x)-(1-x)\log_2(1-x)$ accounts for a decrease in the final secure key rate due to the QBER, which again consists of two contributions: firstly, imperfect polarization measurements, which we assume to occur with a probability of $e^\textrm{pol}=0.004$, estimated from the worst 200GHz WDM visibility value we measured (see Fig.\,\ref{fig:vis}); and secondly, half of all $CC^\textrm{acc}$ (the other half yields clicks in accordance with the quantum state).
The source can only be operated at one certain $P$, but the $t_\textrm{CC}$ can be set differently for each wavelength channel.
Therefore, for each $P$, we sum over all key rates which are individually maximized with regard to $t_\textrm{CC}$, and then sum over all possible channels.\\
In our calculation, we assume a large raw key and asymmetric random basis choice\,\cite{Lo2005}, which are fair assumptions for our high key-rate scenario:
Deploying e.g. Pockels cells switching the measurement basis with a probability of $10^{-3}$ per detected photon would still yield QBER estimates with a confidence better than $\pm$0.1\% while introducing negligible losses. This is assuming 400\,mW pump power with 100\,GHz WDM spacings and data collection for at least 2 minutes of measurement.
This allows us to calculate key rates under conditions of perfect error correction and privacy amplification\,\cite{Neumann2021b}.\\
Additionally, we assumed a maximum of 2\% deadtime-induced loss at 200\,MHz detector count rate.\\

\bibliographystyle{plain}

\end{document}